\documentclass[prd,aps,floats,preprint,preprintnumbers,nofootinbib]{revtex4}
\usepackage{amsmath,amssymb,mathrsfs}
\usepackage{graphicx}
\usepackage{feynmf}
\usepackage{graphics}
\usepackage{amsmath,amscd}
\usepackage{epsfig}
\usepackage{dsfont}
\usepackage{latexsym,oldgerm}
\usepackage{color}
\def\beq{\begin{equation}}
\def\eeq{\end{equation}}
\def\bea{\begin{eqnarray}}
\def\eea{\end{eqnarray}}

\newcommand{\A}{\mathcal{A}_{\theta}}

\newcommand{\M}{\mathcal{M}}

\newcommand{\dx}{{\rm d}}
\newcommand{\e}{{\rm e}}

\newcommand{\I}{\mathds{1}}
\newcommand{\R}{\mathbb{R}}

\newcommand{\Pg}{\mathscr{P}^\uparrow_+}

\newcommand{\F}{\mathscr{F}_\theta}

\begin{document}
\small
\preprint{SU-4252-914\vspace{1cm}} \setlength{\unitlength}{1mm}
\title{Covariant Quantum Fields on Noncommutative Spacetimes
\vspace{0.1cm}}
\author{ A. P.
Balachandran$^a$}\thanks{bal@phy.syr.edu}\author{A. Ibort$^b$}\thanks{albertoi@math.uc3m.es}\author{G. Marmo$^c$}\thanks{marmo@na.infn.it} \author{M. Martone$^{a,c}$}\thanks{mcmarton@syr.edu}
\affiliation{$^{a}$Department of Physics, Syracuse University, Syracuse, NY
13244-1130, USA\\
$^{b}$Departamento de Matem\'aticas, Universidad Carlos III de Madrid, 28911 Legan\'es, 
Madrid, Spain\\
$^{c}$Dipartimento di Scienze Fisiche, University of Napoli and INFN, Via Cinthia I-80126 Napoli, Italy}
\begin{abstract}
A spinless covariant field $\varphi$ on Minkowski spacetime $\M^{d+1}$ obeys the relation $U(a,\Lambda)\varphi(x)U(a,\Lambda)^{-1}=\varphi(\Lambda x+a)$ where $(a,\Lambda)$ is an element of the Poincar\'e group $\Pg$ and $U:(a,\Lambda)\to U(a,\Lambda)$ is its unitary representation on quantum vector states. It expresses the fact that Poincar\'e transformations are being unitary implemented. It has a classical analogy where field covariance shows that Poincar\'e transformations are canonically implemented. Covariance is self-reproducing: products of covariant fields are covariant. We recall these properties and use them to formulate the notion of covariant quantum fields on noncommutative spacetimes. In this way all our earlier results on dressing, statistics, etc. for Moyal spacetimes are derived transparently. For the Voros algebra, covariance and the $*$-operation are in conflict so that there are no covariant Voros fields compatible with $^*$, a result we found earlier. The notion of Drinfel'd twist underlying much of the preceding discussion is extended to discrete abelian and nonabelian groups such as the mapping class groups of topological geons. For twists involving nonabelian groups the emergent spacetimes are nonassociative.
\vspace{0.2cm}
\end{abstract}
\maketitle
\section{INTRODUCTION: Poincar\'e covariance on commutative spacetimes}

The Poincar\'e group $\mathscr{P}$ acts on Minkowski space $\M^{d+1}$ by transforming its coordinates (or coordinate functions), $x=(x_\mu)\ {\rm to}\ \Lambda x+a$
\beq\label{Co1}
(a,\Lambda)\in\mathscr{P}:\quad (a,\Lambda)x=\Lambda x+a\quad.
\eeq

If the spacetime algebra $\mathcal{A}_0(\M^{d+1})$ associated with $\M^{d+1}$ is commutative, and $\varphi$ is a quantum relativistic scalar field on $\M^{d+1}$, we require that there exists a unitary representation
\beq
U:(a,\Lambda)\to U(a,\Lambda)
\eeq
on the Hilbert space $\mathcal{H}$ of states vectors such that 
\beq\label{Co2}
U(a,\Lambda)\varphi(x)U(a,\Lambda)^{-1}=\varphi\big((a,\Lambda)x\big)\quad.
\eeq

There are similar requirements on relativistic fields of all spins. They express the requirement that the spacetime transformations (\ref{Co1}) can be unitarily implemented in quantum theory. It is analogous to the requirement in nonrelativistic quantum mechanics that infinitesimal spatial rotations are to be implemented by the (self-adjoint) angular momentum operators. 

A field $\varphi$ fulfilling (\ref{Co2}) is said to be a ``covariant field'' and the condition in (\ref{Co2}) is the covariance condition. We call it ``primitive'' as we later extend it to products of fields. 

We can write (\ref{Co2}) in the equivalent form 
\beq\label{Co3}
U(a,\Lambda)\varphi\big((a,\Lambda)^{-1}x\big)U(a,\Lambda)=\varphi(x)
\eeq

Now in this form, covariance can be readly understood in terms of the coproduct on the Poincar\'e group. Thus 
\beq
\varphi\in L(\mathcal{H})\otimes S(\M^{d+1})
\eeq
where $L(\mathcal{H})$ are linear operators on $\mathcal{H}$ and $S(\M^{d+1})$ are distributions on $\M^{d+1}$. There is an action of $\mathscr{P}$ on both, that on $L(\mathcal{H})$ being the adjoint action ${\rm Ad}U(a,\Lambda)$ of $U(a,\Lambda)$,
\beq
{\rm Ad}U(a,\Lambda)\varphi=U(a,\Lambda)\varphi U(a,\Lambda)^{-1}
\eeq
and that on $S(\M^{d+1})$ being
\beq
\alpha\to(a,\Lambda)\triangleright\alpha,\quad\big[(a,\Lambda)\alpha\big](x)=\alpha\big((a,\Lambda)^{-1}x\big),\quad\alpha\in S(\M^{d+1})\quad.
\eeq

We call the latter action as $V$.

Now the coproduct on $\Pg$ for commutative spacetimes is $\Delta_0$, where
\beq
\Delta_0\big((a,\Lambda)\big)=(a,\Lambda)\otimes(a,\Lambda)\quad.
\eeq
Then by (\ref{Co3})
\beq\label{Co4}
({\rm Ad}U\otimes V)\Delta_0\big((a,\Lambda)\big)\varphi=\varphi\quad.
\eeq

We will have occasion to use both the versions (\ref{Co2}) and (\ref{Co3},\ref{Co4}) of covariance.

\section{Covariance for products: Commutative Spacetimes}

We saw in the previous section that for a single field, covariance ties together spacetime transformations  and its implementation on the quantum Hilbert space. Products of fields bring in new features which although present for commutative spacetimes, assume prominence on quantum spacetimes. We now briefly examine these features in the former case

\subsection{{\it Tensor Products}}

Consider
\beq
\varphi(x_1)\varphi(x_2)...\varphi(x_N)\quad.
\eeq

This can be understood as the element $\varphi\otimes\varphi...\otimes\varphi$ belonging to $L(\mathcal{H})\otimes\big(S(\M^{d+1})\otimes S(\M^{d+1})\otimes...\otimes S(\M^{d+1})\big)$
evaluated at $x_1,x_2,...,x_N$
\beq
\varphi\otimes\varphi\otimes...\otimes\varphi\in L(\mathcal{H})\otimes\big(S(\M^{d+1})\big)^{\otimes N},\quad(\varphi\otimes\varphi\otimes...\otimes\varphi)(x_1,x_2,...,x_N)=\varphi(x_1)\varphi(x_2)...\varphi(x_N)\quad.
\eeq

Note that tensoring refers only to $S(\M^{d+1})$, there is no tensoring involving $L(\mathcal{H})$. There is only one Hilbert space $\mathcal{H}$ which for free particles is the Fock space and $U(a,\Lambda)$ acts by conjugation on the L.H.S. for all $N$.

But that is not the case for $S(\M^{d+1})^{\otimes N}$. The Poincar\'e group acts on it by the coproduct
\beq\label{Co71}
(\underbrace{\I\otimes\I\otimes...\otimes\I\otimes\Delta_0}_{ N-1})(\underbrace{\I\otimes\I\otimes...\otimes\I\otimes\Delta_0}_{N-2})...\Delta_0
\eeq
of $(a,\Lambda)$. Thus
\beq\label{Co72}
{\rm (\ref{Co71})\ on\ }(a,\Lambda)=(a,\Lambda)\otimes(a,\Lambda)\otimes...\otimes(a,\Lambda)
\eeq
and
\beq
\Big({\rm (\ref{Co71})\ on\ }(a,\Lambda)\triangleright\varphi^{\otimes N}\Big)(x_1,x_2,...,x_N)=\varphi^{\otimes N}\big((a,\Lambda)^{-1}x_1,(a,\Lambda)^{-1}x_2,...,(a,\Lambda)^{-1}x_N\big)\quad.
\eeq

Covariance is now the demand
\beq
U(a,\Lambda)\Big(\varphi^{\otimes N}\big((a,\Lambda)^{-1}x_1,(a,\Lambda)^{-1}x_2,...,(a,\Lambda)^{-1}x_N\big)\Big)U(a,\Lambda)^{-1}=\varphi^{\otimes N}(x_1,x_2,...,x_N)\quad.
\eeq
It is evidently fulfilled for the coproduct (\ref{Co71}) if the primitive covariance (\ref{Co2},\ref{Co3}) is fulfilled. 

For free fields (or in and out-fields), covariance can be verified in a different manner. Thus for a free real scalar field $\varphi$ of mass $m$, we have
\bea\label{Co81}
&\varphi=\int\dx\mu(p)\Big(c^\dag_p\e_p+c_p\e_{-p}\Big)=\varphi^{(-)}+\varphi^{(+)}&\\\nonumber&\e_p(x)=\e^{-ip\cdot x},\quad|p_0|=(\vec{p}^2+m^2)^{\frac{1}{2}},\quad\dx\mu(p)=\frac{\dx^dp}{2|p_0|}&
\eea
where $c_p$, $c^\dag_p$ are the standard annihilation and creation operators, and $\varphi^{(\mp)}$ refer to the annihilation and creation parts of $\varphi$.

Now $\varphi^{(\mp)}$ must separately fulfill the covariance requirement. Let us consider $\varphi^{(-)}$. We have that
\beq\label{Co82}
\varphi^{(-)}(x_1)\varphi^{(-)}(x_2)...\varphi^{(-)}(x_N)|0\rangle=\int\prod_i\dx\mu(p_i)c^\dag_{p_1}c^\dag_{p_2}...c^\dag_{p_N}|0\rangle\e_{p_1}(x_1)\e_{p_2}(x_2)...\e_{p_N}(x_N)
\eeq

Let us first check translations. Let $P_\mu$ be the translation generators on the Hilbert space,
\beq\label{Co91}
[P_\mu,c_p^\dag]=p_\mu c^\dag_p,\quad P_\mu|0\rangle=0
\eeq
and let $\mathcal{P}_\mu=-i\partial_\mu$ be the translation generator on $S(\M^{d+1})$:
\beq\label{Co92}
\mathcal{P}_\mu\e_p=-p_\mu\e_p
\eeq
The coproduct $\Delta_0$ gives for the Lie algebra element $\mathcal{P}_\mu$,
\beq\label{Co93}
\Delta_0(\mathcal{P}_\mu)=\I\otimes\mathcal{P}_\mu+\mathcal{P}_\mu\otimes\I
\eeq
[If $\underline{v}$ is the representation of the Lie algebra of $\Pg$ on functions, and $\hat{P}_\mu$ is the Lie algebra generator in the abstract group $\Pg$ so that $\underline{v}(P_\mu)=\mathcal{P}_\mu$, the L.H.S. here should strictly read $\underline{v}\big(\Delta_0(\hat{P}_\mu)\big)$. So we have simplified the notation in (\ref{Co93}).]
 
It follows that 
\beq
(\I\otimes\I\otimes...\otimes\I\otimes\Delta_0)...\Delta_0(\mathcal{P}_\mu)\e_{p_1}\otimes\e_{p_2}\otimes...\otimes\e_{p_N}=-\sum_i p_{i\mu}\e_{p_1}\otimes\e_{p_2}\otimes...\otimes\e_{p_N}
\eeq

Covariance for translations is the requirement
\beq\label{Co101}
P_\mu c^\dag_{p_1}c^\dag_{p_2}...c^\dag_{p_N}|0\rangle\e_{p_1}\otimes\e_{p_2}\otimes...\otimes\e_{p_N}+c^\dag_{p_1}c^\dag_{p_2}...c^\dag_{p_N}|0\rangle\big(-\sum_ip_{i\mu}\big)\e_{p_1}\otimes\e_{p_2}\otimes...\otimes\e_{p_N}=0
\eeq
which is clearly fulfilled.

Next consider Lorentz transformations. A Lorentz transformation $\Lambda$ acts on $\e_p$ according to
\beq
(\Lambda\e_p)(x)=\e_p(\Lambda^{-1}x)=\e_{\Lambda p}(x)
\eeq
or $\Lambda\e_p=\e_{\Lambda p}$.

For Lorentz transformations $\Lambda$, covariance is thus the identity
\beq
\int\prod_i\dx\mu(p_i)c^\dag_{\Lambda p_1}c^\dag_{\Lambda p_2}...c^\dag_{\Lambda p_N}|0\rangle\e_{\Lambda p_1}\otimes\e_{\Lambda p_2}...\otimes\e_{\Lambda p_N}=\int\prod_i\dx\mu(p_i)c^\dag_{p_1}c^\dag_{p_2}...c^\dag_{p_N}|0\rangle\e_{p_1}\otimes\e_{p_2}...\otimes\e_{p_N}
\eeq
which is true because of the Lorentz invariance of the measure:
\beq
\dx\mu(\Lambda^{-1}p_i)=\dx\mu(p_i)\quad.
\eeq

\section{Quantum Statistics: the Schur-Weyl Duality}

The permutation group $S_N$ and its irreducible representations govern statistics of $N$-particle state vectors on commutative spacetimes for $d\geq3$. We consider only such $d$.

By axioms of quantum theory, the $N$-particle observables must commute with the action of $S_N$ so that the action of observation does not affect particle identity. In particular the action of the symmetry group must commute with the action of $S_N$.

If that is the case, we can consistently work with irreducible representations of $S_N$. 

In (\ref{Co82}), $(a,\Lambda)$ acts on $\e_{p_1}\otimes\e_{p_2}\otimes...\otimes\e_{p_N}$ via the coproduct (\ref{Co71}). This action commutes with the action of $S_N$ if $S_N$ acts by permuting $p_i$. Thus we can work with irreducible representation of $S_N$.

In particular we can work with bosons and fermions by totally symmetrising or antisymmetrising $\otimes\e_{p_i}$. In the former case $c^\dag_{p_i}$ can be taken to commute (their anticommutators do not contribute to (\ref{Co82})) and for the latter they anticommute.

The important point here is that the group algebras $\mathbb{C}\mathscr{P}$ and $\mathbb{C}S_N$ are commutants of each other in their action on $N$-particle states.

\subsection{{\it The Double Commutant Theorem and the Schur-Weyl Duality}}

A result of this sort is familiar to particle physicists in case the symmetry group is $U(k)$. Here $U(k)$ can be the $k$-flavour symmetry group. It acts on $\mathbb{C}^k$. Then to reduce the representation of $U(k)$ on $\mathbb{C}^{k\otimes N}$, we use the fact that $\mathbb{C}S_N$ commutes with $\mathbb{C}U(k)$. That lets us use Young tableaux methods.

It is in fact the case that $\mathbb{C}U(k)$ and $\mathbb{C}S_N$ exhaust the commutants of each other. This result and the Young tableaux methods are part of the contents of Schur-Weyl duality \cite{Schur1,Schur2}.

So we are working with aspects of an infinite-dimensional analogue of this duality for a noncompact symmetry group $\Pg$ when we remark that $\mathbb{C}\Pg$ and $\mathbb{C}S_N$ mutually commute.

\subsection{{\it A Presentation of $S_N$}}

Let us imagine that $S_N$ acts by transforming $N$ objects numbered from 1 to $N$ and let $\tau_{ij}$ denote the transformation of objects $i$ and $j$. Then $S_N$ has the presentation
\beq\label{Co141}
S_N=\langle\tau_{i,i+1}:i\in[1,2,...,N-1],\tau^2_{i,i+1}=\I,\tau_{i,i+1}\tau_{i+1,i+2}\tau_{i,i+1}=\tau_{i+1,i+2}\tau_{i,i+1}\tau_{i+1,i+2}\rangle
\eeq

The $N$ objects were introduced here for concreteness. The abstract $S_N$ group is given just by (\ref{Co141}).

We will have use of this presentation later.

\subsection{{\it Multiplication Map and Self-Reproduction}}

The multiplication map involves products of fields at the {\it same} point and hence the algebra of the underlying manifold. It is not the same as the tensor product which involves products of fields at {\it different} points.

There is a further property of $\varphi$, involving now the multiplication map, which is easily understood on commutative spacetimes. It has much importance for both commutative and noncommutative spacetimes. It is the property of self-reproduction. Let us first understand this property for $C^\infty(\M)$, the set of smooth functions on a manifold $\M$. If $\alpha:p\to \alpha p,\ p\in\M$, is a diffeomorphism of $\M$, it acts on $f\in C^\infty(\M)$ by pull-back:
\beq\label{Co151}
(\alpha^*f)(p)=f(\alpha p)\quad.
\eeq

But $C^\infty(\M)$ has a further property, routinely used in differential geometry: $C^\infty(\M)$ is closed under point-wise multiplication:

If $f_1,f_2\in C^\infty(\M)$, then
\beq\label{Co152}
f_1f_2\in C^\infty(\M)
\eeq
where
\beq\label{Co153}
\big(f_1f_2\big)(p)=f_1(p)f_2(p)\quad.
\eeq

This property is very important for noncommutative geometry: the completion of this algebra under the supremum norm gives the commutative algebra of $C^0(\M)$, a commutative $C^*$-algebra. By the Gel'fand-Naimark theorem \cite{Landi, Varilly} it encodes the topology of $\M$.

Now by (\ref{Co151}) and (\ref{Co152}), we see that multiplication of functions preserves transformation under diffeos. This simple property gets generalised to covariant quantum field thus:

{\it The pointwise product of covariant quantum fields is covariant.}

That means in particular that
\beq
U(a,\Lambda)\varphi^2\big((a,\Lambda)^{-1}x\big)U(a,\Lambda)^{-1}=\varphi^2(x)\quad.
\eeq
This result is obviously true modulo renormalization problems. It is at the basis of writing invariant interactions in quantum field theories  on $\mathcal{A}_0(\M^{d+1})$.

Note that generally we require covariance of the product of any two covariant fields, distinct or the same.

\subsection{{\it The $*$-covariance}}

In quantum field theories on $\mathcal{A}_0(\M^{d+1})$, another routine requirement is that covariance and the $*$- or the adjoint operation be compatible. Thus if $\psi$ is a covariant complex field,
\beq
U(a,\Lambda)\psi\big((a,\Lambda)^{-1}x\big)U(a,\Lambda)^{-1}=\psi(x)\quad,
\eeq
we require that $\psi^\dag$ is also a covariant complex field. That is fulfilled if $U(a,\Lambda)$ is unitary. 

{\it Thus $*$-covariance is linked to unitarity of time-evolution and the S-matrix and many more physical requirements}.

\subsection{{\it Summary: Covariance Requirements}}

Here is a brief summary of our covariance requirements on quantum fields for commutative spacetimes (ignoring the possibility of parastatistics of order 2 or more):
{\it A quantum field should be $*$- covariant with commutation or anti-commutation relations (symmetrisation postulates) compatible with $*$-covariance.}

\section{Covariance on the Moyal Plane}

The Moyal plane $\A(\M^{d+1})$ is the algebra of smooth functions on $\M^{d+1}$ with the product
\beq\label{Co181}
m_\theta(\alpha\otimes\beta)=m_0\F(\alpha\otimes\beta),\quad \alpha,\beta\in\A(\M^{d+1}),\quad\F=\e^{\frac{i}{2}\partial_\mu\otimes\theta_{\mu\nu}\partial_\nu}
\eeq
where $m_0$ is the point-wise product:
\beq
m_0(\gamma\otimes\delta)(x)=\gamma(x)\delta(x),\quad\gamma,\delta\in\mathcal{A}_0(\M^{d+1})\quad.
\eeq

The Poincar\'e group $\mathscr{P}$ acts on smooth functions $\alpha$ on $\M^{d+1}$ by pull-back as before:
\beq\label{Co182}
\mathscr{P}\ni(a,\Lambda):\alpha\to(a,\Lambda)\alpha,\quad\big((a,\Lambda)\alpha\big)(x)=\alpha\big((a,\Lambda)^{-1}x\big)
\eeq

It is by now well-known \cite{Chaichian, Aschieri1,Aschieri2} that this action extends to the algebra $\A(\M^{d+1})$ compatibly with the product $m_\theta$ only if the coproduct on $\mathscr{P}$ is twisted. The twisted coproduct $\Delta_\theta$ on $\mathscr{P}$ is 
\beq\label{Co183}
\Delta_\theta(g)=F^{-1}_\theta(g\otimes g)F_\theta,\quad F_\theta=\e^{-\frac{i}{2}\hat{P}_\mu\otimes\theta_{\mu\nu}\hat{P_\nu}}={\rm Drinfel'd\ twist}
\eeq

Here $\hat{P}_\mu$ is as before the translation generator in $\mathscr{P}$ with representatives $\mathcal{P}_\mu=-i\partial_\mu$ and $P_\mu$ on functions and $L(\mathcal{H})$ respectively.

Equation (\ref{Co183}) is the starting point for further considerations.

Let $\varphi_\theta$ be the twisted analogue of the field $\varphi$ of section 2. Also let $U_\theta$ be the unitary operator implementing $\mathscr{P}$ in $L(\mathcal{H})$. Covariance then is the requirement
\beq\label{Co191}
U_\theta(a,\Lambda)\varphi_\theta\big((a,\Lambda)^{-1}x\big)U_\theta(a,\Lambda)^{-1}=\varphi_\theta(x)
\eeq
and its multifield generalisation, while compatibility with $^*$ or unitarity requires that $\varphi_\theta^\dag$ is also covariant. There is also one further requirement, namely compatibility with symmetrisation postulate. 

The analysis of these requirements becomes transparent on working with the mode expansion of $\varphi_\theta$ which is assumed to exist:
\beq\label{Co192}
\varphi_\theta=\int\dx\mu(p)\big[a^\dag_p\e_p+a_p\e_{-p}\big]=\varphi_\theta^{(-)}+\varphi_\theta^{(+)},\quad\dx\mu(p)=\frac{\dx^dp}{2|p_0|}\quad.
\eeq
The expansion can refer to in- , out- or free fields. 

We also assume the existence of vacuum $|0\rangle$:
\beq\label{Co201}
a_p|0\rangle=0,\ \forall p\quad.
\eeq
\subsection{{\it The Primitive Covariance of a Single Field}}

We are here referring to (\ref{Co192}). It requires that
\beq\label{Co202}
U_\theta(a,\Lambda)a^\dag_pU_\theta(a,\Lambda)^{-1}=a^\dag_{\Lambda p},\quad U_\theta(a,\Lambda)a_pU_\theta(a,\Lambda)^{-1}=a_{\Lambda p}
\eeq
A particular consequence of (\ref{Co201},\ref{Co202}) is that single particle states transform for all $\theta$ in the same manner or assuming that $U_\theta(a,\Lambda)|0\rangle=|0\rangle$:
\beq
U_\theta(a,\Lambda)a^\dag_p|0\rangle=a^\dag_{\Lambda p}|0\rangle
\eeq
New physics can be expected only in multi-particle sectors.

\subsection{{\it Covariance in Multi-Particle Sectors}}

On the Moyal plane, multi-particle wave functions $\e_{p_1}\otimes\e_{p_2}\otimes...\otimes\e_{p_N}$ transform under $\mathscr{P}$ with the twisted coproduct. This affects the properties of $a_p$, $a^\dag_p$ in a $\theta_{\mu\nu}$-dependent manner.

Let us focus on the two-particle sector:
\beq\label{Co211}
\int\prod_i\dx\mu(p_i)a^\dag_{p_1}a^\dag_{p_2}|0\rangle\e_{p_1}\otimes\e_{p_2}
\eeq
Since translations act in the usual way on $\e_{p_1}\otimes\e_{p_2}$,
\beq\label{Co212}
\Delta_\theta(\mathcal{P}_\mu)\e_{p_1}\otimes\e_{p_2}=(\I\otimes\mathcal{P}_\mu+\mathcal{P}_\mu\otimes\I)\e_{p_1}\otimes\e_{p_2}=-(\sum_ip_{i\mu})\e_{p_1}\otimes\e_{p_2}
\eeq
translational covariance requires the standard transformation of $a^\dag_{p_i}$:
\beq\label{Co213}
[P^\theta_\mu,a^\dag_p]=p_\mu a^\dag_p\quad,
\eeq
$P^\theta_\mu$ is the possibly $\theta$ dependent translation generator.

Lorentz transformations are more interesting. We have that
\beq\label{Co221}
\Delta_\theta(\Lambda)\triangleright\e_{p_1}\otimes\e_{p_2}=\F^{-1}(\Lambda\otimes\Lambda)\F\e_{p_1}\otimes\e_{p_2}=\e^{\frac{i}{2}(\Lambda p_1)\wedge(\Lambda p_2)}\e^{-\frac{i}{2}p_1\wedge p_2}\e_{\Lambda p_1}\otimes\e_{\Lambda p_2}\quad.
\eeq
(We do not consider the anti-unitary time-reversal in what follows.) Covariance thus requires that
\beq\label{Co222}
\int\prod_i\dx\mu(p_i)U_\theta(\Lambda)a^\dag_{p_1}a^\dag_{p_2}|0\rangle\e^{\frac{i}{2}(\Lambda p_1)\wedge(\Lambda p_2)}\e^{-\frac{i}{2}p_1\wedge p_2}\e_{\Lambda p_1}\otimes\e_{\Lambda p_2}=\int\prod_i\dx\mu(p_i)a^\dag_{p_1}a^\dag_{p_2}|0\rangle\e_{\Lambda p_1}\otimes\e_{\Lambda p_2}
\eeq

\subsection{{\it The Dressing Transformation}}

We can solve this requirement, as well as (\ref{Co213}), by writing $a^\dag_p$ in terms of the $c^\dag_p$ and $P_\mu$:
\beq\label{Co223}
a^\dag_p=c^\dag_p\e^{\frac{i}{2}p\wedge P}
\eeq
and setting
\beq
U_\theta(a,\Lambda)=U_0(a,\Lambda)=U(a,\Lambda)\quad.
\eeq
The adjoint of (\ref{Co223}) is
\beq\label{Co231}
a_p=\e^{-\frac{i}{2}p\wedge P}c_p=c_p\e^{-\frac{i}{2}p\wedge P}\quad,
\eeq
where the equality in the last step uses the anti-symmetry of $\theta_{\mu\nu}$.

As we can twist $c_p$ on left {\it or} on right, we can write $\varphi_\theta$ as a twist applied to $\varphi_0\equiv\varphi$:
\beq\label{Co232}
\varphi_\theta=\varphi_0\e^{-\frac{1}{2}\overleftarrow{\partial}\wedge P}
\eeq
The transformation $\varphi_0\to\varphi_\theta$ is an example of a dressing transformation. It was first introduced in the context of integrable models by Grosse \cite{Grosse} and by Faddeev and Zamalodichkov \cite{dress1,dress2}.

It is important to note that (\ref{Co232}) {\it is well-defined for a fully interacting Heisenberg field $\varphi_0$ if $P_\mu$ stands for the total four momentum of the interacting theory}. In that case $\varphi_\theta$ is the twisted Heisenberg field. 

We can now check that 
\beq\label{Co241}
U(a,\Lambda)\varphi_\theta(x_1)\varphi_\theta(x_2)...\varphi_\theta(x_N)U(a,\Lambda)^{-1}|0\rangle=\varphi_\theta\big((a,\Lambda)x_1\big)\varphi_\theta\big((a,\Lambda)x_2\big)...\varphi_\theta\big((a,\Lambda)x_N\big)|0\rangle
\eeq
with a similar equation for the vacuum $\langle0|$ put on the left. Since vacuum is a cyclic vector, we can then be convinced that (\ref{Co232}) fully solves the problem of constructing a covariant quantum field on the Moyal plane at the multi-field level as well. 

A particular implication of (\ref{Co241}) is that 
\beq
U_\theta(a,\Lambda)=U(a,\Lambda)=U_0(a,\Lambda)\quad.
\eeq
Its expression in terms of in-, out- or free fields looks the same as in the commutative case. It has no $\theta_{\mu\nu}$- dependence. 

\subsection{{\it Symmetrization and Covariance}}

We will now show that the dressing transformations (\ref{Co223},\ref{Co231}-\ref{Co232}) are exactly what we need to be compatible with appropriate symmetrisation postulates. 

At the level of the particle dynamics (functions on $\M^{d+1}$ and their tensor products), it is known that for the coproduct $\Delta_\theta$, symmetrisation and anti-symmetrisation should be based on the twisted flip operator
\bea\label{Co251}
\tau_\theta=\F^{-1}\tau_0\F\\\label{Co252}
\tau_0\alpha\otimes\beta:=\beta\otimes\alpha
\eea
where $\alpha,\beta$ are single particle wave functions.

As defined, $\tau_0$ and $\tau_\theta$ act on two-particle wave functions and generate $S_2$ since 
\beq
\tau^2_0=\I\quad\Rightarrow\quad\tau^2_\theta=\I\quad.
\eeq
But soon we will generalise them to $N$-particles to get $S_N$.

Thus twisted  bosons (fermions) have the two-particle plane wave states
\beq
\e_{p_1}\otimes_{S_\theta}\e_{p_2}=\frac{\I\pm\tau_\theta}{2}\e_{p_1}\otimes\e_{p_2}\quad.
\eeq

Let us focus on $S_\theta$:
\bea
\e_{p_1}\otimes_{S_\theta}\e_{p_2}&=&\frac{1}{2}\left[\e_{p_1}\otimes\e_{p_2}+\F^{-2}\e_{p_2}\otimes\e_{p_1}\right]\\
&=&\frac{1}{2}[\e_{p_1}\otimes\e_{p_2}+\e^{ip_2\wedge p_1}\e_{p_2}\otimes\e_{p_1}]\\
&=&\e^{ip_2\wedge p_1}\e_{p_2}\otimes_{S_\theta}\e_{p_1}
\eea
This gives
\bea
&&\int\prod_{i=1}^2\dx\mu(p_i)a^\dag_{p_1}a^\dag_{p_2}|0\rangle\e_{p_1}\otimes_{S_\theta}\e_{p_2}\\
&&\qquad=\int\prod_{i=1}^2\dx\mu(p_i)a^\dag_{p_1}a^\dag_{p_2}|0\rangle\e^{ip_2\wedge p_1}\e_{p_2}\otimes_{S_\theta}\e_{p_1}\\
&&\qquad=\int\prod_{i=1}^2\dx\mu(p_i)\big(\e^{ip_1\wedge p_2}a^\dag_{p_2}a^\dag_{p_1}\big)|0\rangle\e_{p_1}\otimes_{S_\theta}\e_{p_2}
\eea

Thus we require that
\beq
a^\dag_{p_1}a^\dag_{p_2}=\e^{ip_1\wedge p_2}a^\dag_{p_2}a^\dag_{p_1}
\eeq
which is fulfilled by (\ref{Co223}).

We can extend this demonstration regarding the consistency of the twist to multinomials in $a^\dag$'s and $a$'s. The necessary tools are in \cite{sasha}. We just note one point. In the $N$-particle sector, call $\F^{ij}$ the Drinfel'd twist (\ref{Co181}) where in $\partial_\mu\otimes\partial_\nu$, $\partial_\mu$ acts on the $i^{\rm th}$ and $\partial_\nu$ on the $j^{\rm th}$ factor in the tensor product.

Define
\beq
\tau^{ij}_\theta=\F^{-1}\tau_0^{ij}\F=\F^{-2}\tau_0^{ij}
\eeq
where $\tau^{ij}_0$ flips the entries of an $N$-fold tensor product by flipping the $i^{\rm th}$ and $j^{\rm th}$ entries as in (\ref{Co252}). Then
\beq
\left(\tau_0^{ij}\right)^2=\I
\eeq
which is obvious and
\beq
\tau^{i,i+1}_\theta\tau_\theta^{i+1,i+2}\tau^{i,i+1}_\theta=\tau^{i+1,i+2}_\theta\tau_\theta^{i,i+1}\tau^{i+1,i+2}_\theta
\eeq
which is not obvious. It follows from (\ref{Co141}) that $\tau_\theta^{i,i+1}$'s generate $S_N$ in this sector. 

One can check that the Poincar\'e group action with the twisted coproduct commutes with this action of $S_N$.

\subsection{{\it $*$-Covariance}}

Covariance requirements on the Moyal plane has led us to the dressed field (\ref{Co232}). We now require it to be compatible with the $*$-operation. That is if $\varphi^*_0=\varphi_0$, we want that $\varphi^*_\theta=\varphi_\theta$. Now
\beq
\varphi_\theta^*=\e^{-\frac{1}{2}\partial\wedge P}\varphi_0
\eeq
where $\partial_\mu$ acts just on $\varphi_0$, $P_\nu$ acts on $\varphi_0$ and all that may follow. But since $P_\nu$ acting on $\varphi_0$ is $-i\partial_\nu\varphi_0$ and $\partial\wedge\partial=0$, we see that
\beq
\varphi^*_\theta=\varphi^*_0\e^{-\frac{1}{2}\overleftarrow{\partial}\wedge P}\quad.
\eeq
So the dressing transformations preserves $*$-covariance. The antisymmetry of $\theta$ plays a role in this process.

We can also understand these statements from (\ref{Co223}). That gives
\beq
a_p=\e^{-\frac{i}{2}p\wedge P}c_p=c_p\e^{-\frac{i}{2}p\wedge P}
\eeq
since $p\wedge p=0$. So {\it we can twist both creation and annihilation operators on the same side because $\theta$ is antisymmetric}. It is only because of this that we can get the twisted quantum Heisenberg field (\ref{Co232}). The importance of its existence has been emphasised before.

\section{Moyal vs Voros}

The Voros plane $\mathcal{A}_\theta^V(\M^{d+1})$ is the algebra of functions on $\M^{d+1}$ with the star product
\beq
\alpha\star_V\beta=m_0(\F^V\alpha\otimes\beta),\quad\F^V=\e^{\frac{i}{2}\partial_\mu\otimes(\theta_{\mu\nu}-iS_{\mu\nu})\partial_\nu}
\eeq
where $S_{\mu\nu}=S_{\nu\mu}$ defines a constant real {\it symmetric} matrix. The matrix $\theta$ fixes $S$, we will see how this happens for general $d$ later.

But for $d=1$, this determination is easy to describe. For $d=1$, $\theta_{\mu\nu}=\hat{\theta}\epsilon_{\mu\nu}$, $\epsilon_{12}=-\epsilon_{21}=1$, $\epsilon_{11}=\epsilon_{22}=0$ and then $S_{\mu\nu}=\hat{\theta}\delta_{\mu\nu}$. So for $d=1$,
\beq\label{Co311}
\F^V=\e^{\frac{i}{2}\partial_\mu\otimes\theta_{\mu\nu}\partial_\nu+\hat{\theta}\partial_\mu\cdot\partial_\mu}
\eeq
where $\partial_\mu\cdot\partial_\mu$ is defined using the {\it Euclidean} scalar product:
\beq
\partial_\mu\cdot\partial_\mu:=\sum^1_{i=0}\partial_\nu\partial_\nu
\eeq

Let us first consider $d=1$.

On plane waves $\e_p$ ($\e_p(x)=\e^{-ip\cdot x}$), the Voros product is
\beq\label{Co312}
\e_p\star_V\e_q=\e^{-\frac{1}{2}\hat{\theta} p\cdot q}\e^{-\frac{i}{2}p\wedge q}\e_{p+q}
\eeq
where $p\cdot q$ is also defined using the Euclidean scalar product:
\beq\label{Co313}
p\cdot q=\sum_{\nu=0}^1p_\nu q_\nu\quad.
\eeq

It is well-known that $\A^\M$ and $\A^V$ are $*$-isomorphic algebras. Thus let
\beq
{\rm T}:\A^\M\to\A^V,\qquad {\rm T}\e_p=\e^{-\frac{1}{4}\hat{\theta}p^2}\e_p
\eeq
Then a simple calculation shows that
\beq
{\rm T}(\e_p\star_\M\e_q)=({\rm T}\e_p)\star_V({\rm T}\e_q),\quad{\rm T}(\bar{\e}_p)=\overline{({\rm T}\e_p)}
\eeq
where bar denotes complex conjugation and $\star_\M$ denotes the Moyal product. (We denoted it previously as just $\star$.)

The $*$-isomorphism of $\A^{\M,V}$ may suggest that quantum field theories are not sensitive to which algebra we use. But that is not the case. Thus we should require that the twisted (dressed) in- (out-) creation and annihilation operators on $\A^V$ are adjoints of each other for $*$-covariance. But this imposition spoils the possibility of constructing Heisenberg fields.

On the other hand, a naive construction of the dressed Heisenberg field is incompatible with the adjoint operation: such a dressing applied to a self-adjiont field is not self-adjoint.

These results have been discussed before \cite{Mario,Mario3}. Here we recall the proofs.

Let us first assume that the Voros $\star$ also admits twisted creation-annihilatin operators and associated (in-, out-, or free-) field $\varphi_{\theta,V}$ as in (\ref{Co192}):
\beq\label{Co331}
\varphi_{\theta,V}=\int\dx\mu(p)\left[a^\dag_{p,V}\e_p+a_p\e_{-p}\right]:=\varphi^{(-)}_{\theta,V}+\varphi^{(+)}_{\theta,V}
\eeq

Primitive covariance gives as before
\beq
U(a,\Lambda)a_{p,V}U(a,\Lambda)^\dag=a_{\Lambda p,V}
\eeq
and 
\beq
U(a,\Lambda)a^\dag_{p,V}U(a,\Lambda)^\dag=a^\dag_{\Lambda p,V}
\eeq
where we did not attach a $\theta$ to $U$. 

In the two-particle sector, the coproduct by general principles is
\beq\label{Co3321}
\Delta_{\theta,V}(g)=\mathscr{F}^{-1}_{\theta,V}(g\otimes g)\mathscr{F}_{\theta,V}
\eeq
As $\mathscr{F}_{\theta,V}$ is translationally invariant, the coproduct for $P_\mu$ is not affected by the twist. So we focus on Lorentz transformations.

For Lorentz transformations, (\ref{Co222}) is modified to 
\beq
\left(\int\prod\dx\mu(p_i)U(\Lambda)a^\dag_{p_1,V}a^\dag_{p_2,V}|0\rangle\right)\e^{\frac{i}{2}(\Lambda p_1)\wedge (\Lambda p_2)-\frac{\hat{\theta}}{2}(\Lambda p_1)\cdot(\Lambda p_2)}\e^{-\frac{i}{2}p_1\wedge p_2-\frac{\hat{\theta}}{2}p_1\cdot p_2}\e_{\Lambda p_1}\otimes \e_{\Lambda p_2}
\eeq
giving the dressing equation
\beq\label{Co341}
a^\dag_{p,V}=c^\dag_p\e^{\frac{i}{2}p\wedge P-\frac{\hat{\theta}}{2}p\cdot P}\quad,
\eeq
scalar products being Euclidean.

The adjoint of (\ref{Co341}) is
\beq
a_{p,V}=\e^{-\frac{i}{2}p\wedge P-\frac{\hat{\theta}}{2}p\cdot P}c_p=\e^{\frac{\hat{\theta}}{2}p\cdot p}c_p\e^{-\frac{i}{2}p\wedge P-\frac{\hat{\theta}}{2}p\cdot P}
\eeq
which is not what we get by dressing $c_p$ on the right.

The result is that $\varphi_{\theta,V}$ is not the outcome of dressing $\varphi_{0,V}$ by a single twist. Its parts $\varphi_{\theta,V}^{(\mp)}$ get separate twists. 

But then there is no way to dress a fully interacting Heisenberg field $\Phi_0$ since $\Phi_0$ cannot decomposed into positive and negative frequency parts.

Or else we can declare that the Voros Heisenberg field is
\beq
\Phi_{\theta,V}=\Phi_0\e^{\frac{1}{2}\overleftarrow{\partial}\wedge P+i\frac{\hat{\theta}}{2}\overleftarrow{\partial}\cdot P}
\eeq
But then if $\Phi^\dag_0=\Phi_0$, $\Phi^\dag_{\theta,V}\neq\Phi_{\theta,V}$. Unitarity is spoilt.

It seems that the Voros plane is not suitable for quantum field theories.

If $d\neq1$, say $d=3$, then by a change of coordinates, we can bring it to the form 
\beq
\hat{\theta}_1\epsilon_{ab}+\hat{\theta}_2\epsilon_{a'b'}\quad (a,b\in[0,1],\ a',b'\in[2,3])\quad.
\eeq
The preceding considerations then apply separately to $\hat{\theta}_1\epsilon_{ab}$ and $\hat{\theta}_2\epsilon_{a'b'}$.

\section{Discrete Groups}

Covariance is a notion tied to symmetry group, and in our context especially to spacetime diffeomorphism groups.

A particularly interesting class of such symmetry groups are mapping class groups of manifolds. They are discrete and for spatial hypersurfaces supporting topological geons can be abelian and nonabelian. In this section we recall our discussion of covariant fields for such geon spatial slices from \cite{GeonMario}.

\subsection{{\it Covariant Quantum Fields for Commutative Geons}}

Let $\mathcal{P}$ be a prime three-manifold, and $\R^3\#\mathcal{P}$ the spatial slice where $\#$ denotes connected sum. Spacetime is then $(\R^3\#\mathcal{P})\#\R$.

Let $D^\infty/D^\infty_0$ be the mapping class group of $\R^3\#\mathcal{P}$ where $D^\infty$ is the diffeo group which keeps a point $p$ (``infinity'') of $\R^3\#\mathcal{P}$ and a frame at $p$ fixed and $D^\infty_0$ its identity component. If $\varphi_0$ is a covariant quantum field, primitive covariance requires that
\begin{itemize}
\item[a)] There is a unitary representation
\beq
U:\quad g^\infty\to U(g^\infty),\quad g^\infty\in D^\infty
\eeq
such that
\beq\label{Co371}
U(g^\infty)\varphi_0(p)U(g^\infty)^{-1}=\varphi(g^\infty p)\quad.
\eeq
\end{itemize}

In addition, constraints in gravity theories require that
\begin{itemize}
\item[b)] $\varphi_0(g^{\infty}_0p)=\varphi_0(p),\quad g^\infty_0\in D^\infty_0$.

\end{itemize}

Note that by b), (\ref{Co371}) can be interpreted in terms of a unitary representation of $D^\infty/D_0^\infty$.

For the Poincar\'e group, the twists were all based on the abelian translation group. Likewise, for now we will base our considerations on twists on the maximal compact abelian subgroup
\beq
A=\times_{i=1}^k\mathbb{Z}_{n_i}
\eeq
There is no loss of generality in assuming compactness as non-compact factors like $\mathbb{Z}$ do not enter the twist \cite{GeonMario}.

We now choose suitable basis of functions for $\R^3\#\mathcal{P}$ adapted to $A$.

Pick a Riemannian metric for $\R^3\#\mathcal{P}$. Its volume form defines a Hilbert space $\mathcal{H}$ of functions on $\R^3\#\mathcal{P}$.

Now the unitary irreducible represenation UIRR $m_i\in\mathbb{Z}/(n_i\mathbb{Z}):=\mathbb{Z}_{n_i}$ of $\mathbb{Z}_{n_i}$ is defined by
\beq
\mathbb{Z}_{n_i}\ni\xi=\e^{i\frac{2\pi}{n_i}}\to\xi^{m_i}\quad.
\eeq
So the UIRR's of A are defined by
\beq
\vec{m}=(m_1,...,m_k)
\eeq
where $m_i$ and $m_i+n_i$ are identified.

Since $A$ acts on $\R^3\#\mathcal{P}$ and hence on $\mathcal{H}$, the latter has an orthonormal basis $\{f^{(\pm)}_{\vec{m}}\}$ which carries the UIRR $\vec{m}$ of $A$ and have positive and negative frequencies $\pm|E_{\vec{m}}|$:
\bea
&f^{(\pm)}_{\vec{m}}(g^\infty_0 p)=f^{(\pm)}_{\vec{m}}(p),\quad g^\infty_0\in D^\infty_0&\quad,\\
&f^{(\pm)}_{\vec{m}}(h^{-1}p)=f^{(\pm)}_{\vec{m}}(p)\chi_{\vec{m}}(h),\quad h\in D^\infty&\quad,\\
&i\partial_0f^{(\pm)}_{\vec{m}}=\pm|E_{\vec{m}}|f^{(\pm)}_{\vec{m}}\quad.&
\eea
(We postulate that such $f^{(\pm)}_{\vec{m}}$ exist.)

Here $\chi_{\vec{m}}$ is the character in UIRR $\vec{m}$. Since $\bar{\chi}_{\vec{m}}=\chi_{-\vec{m}}$, we can assume that
\beq
\bar{f}^{(\pm)}_{\vec{m}}=f^{(\mp)}_{-\vec{m}}
\eeq

If $g\in D^\infty$, we can write
\beq
f^{(\pm)}_{\vec{m}}(g^{-1} p)=\sum_{\vec{m}'}f^{(\pm)}_{\vec{m}'}(p)\mathscr{D}_{\vec{m}'\vec{m}}(g)
\eeq
where $\mathscr{D}$ is a unitary representation of $D^\infty$ which restricted to $D^\infty_0$ becomes the trivial representation. 

The untwisted quantum field (in, out or free) has the mode expansion 
\beq
\varphi_0=\sum_{\vec{m}}\big[c_{\vec{m}}f^{(+)}_{\vec{m}}+c^\dag_{\vec{m}}f^{(-)}_{-\vec{m}}\big]\quad.
\eeq
Then since $\bar{\mathscr{D}}_{\vec{m}'\vec{m}}\mathscr{D}_{\vec{n}'\vec{m}}=\delta_{\vec{m}'\vec{n}'}$, commutative covariance translates to the transformation law
\bea
U(g)c_{\vec{m}}U(g)^{-1}=c_{\vec{m}'}\bar{\mathscr{D}}_{\vec{m}',\vec{m}}(g)\quad,\\
U(g)c^\dag_{\vec{m}}U(g)^{-1}=c^\dag_{\vec{m}'}\mathscr{D}_{\vec{m}',\vec{m}}(g)\quad.
\eea

\subsection{{\it Covariant Geon Fields for Abelian Twists}}

This material (just as the preceding material) has been reported elsewhere \cite{GeonMario}. So we will be brief.

Let $\mathbb{P}_{\vec{m}}$ be the projector in the group algebra $\mathbb{C}A$ to the UIRR $\vec{m}$. Then the Drinfel'd twist based on $A$ is
\beq\label{Co411}
F_\theta=\sum_{\vec{m}',\vec{m}}\e^{-\frac{i}{2}m_i\theta_{ij}m'_j}\mathbb{P}_{\vec{m}}\otimes\mathbb{P}_{\vec{m}'},\quad\theta_{ij}=-\theta_{ji}=\frac{4\pi}{n_{ij}},\quad n_{ij}\ {\rm divdes}\ n_i\ {\rm and}\ n_j\quad.
\eeq

The mode expansion of the twisted field $\varphi_\theta$ is 
\beq
\varphi_\theta=\sum_{\vec{m}}[a_{\vec{m}}f^{(+)}_{\vec{m}}+a^\dag_{\vec{m}}f^{(-)}_{-\vec{m}}]
\eeq
In \cite{GeonMario}, we show that the requirements of covariance for multiparticle states, twisted symmetrisation and self-reproduction are all compatible with the following expression for $a_{\vec{m}}$, $a^\dag_{\vec{m}}$:
\bea
a_{\vec{m}}=\sum_{\vec{m}'}c_{\vec{m}}\e^{-\frac{i}{2}m_i\theta_{ij}m'_j}\mathbb{P}_{m'_j}\quad,\\
a^\dag_{\vec{m}}=\sum_{\vec{m}'}c^\dag_{\vec{m}}\e^{-\frac{i}{2}m_i\theta_{ij}m'_j}\mathbb{P}_{m'_j}\quad.
\eea

\subsection{{\it Non-abelian Twsts}}

Twists such as $F_\theta$ based on abelian groups $A$ lead to associative spacetimes. They can be generalised to twists based on nonabelian group algebras. They lead to nonassociative spacetimes \cite{Nonass,GeonMario}.

A brief examination of covariant quantum fields for such twists is contained in \cite{GeonMario}. It requires more elaboration. In particular not only does spacetime become nonassociative, the coproduct on the symmetry group also loses coassociativity: the symmetry algebra becomes quasi-Hopf \cite{Nonass}. The implications of nonassociativity and quasi-Hopf algebras for quantum field theory and phenomenology remain unexplored.

\section{Final Remarks}
Many papers have been written regarding quantum fields on the Moyal and similar algebras \cite{review,BAP,BP} and on geon spacetimes as well. (See \cite{GeonMario} for references.) In much of this work, quantum fields were constructed using the dressing transformation. This paper systematically clarifies the conceptual basis behind this transformation: it is just covariance. The latter in essence means that symmetry transformations on spacetime and associated structures like suitable symmetrisation postulates of particle wave functions are implementable in the quantum Hilbert space. In classical theory, the analogous requirement would be the canonical implementability of symmetry transformations.

From this point of view, it is clear that covariance and dressing are sensible ideas to construct quantum fields on spacetimes based on Drinfel'd twists. 

Noncommutative spacetimes lead to theories which are acausual and violate Poincar\'e invariance in scattering processes. They violate CPT as well and can lead to Pauli-forbidden transitions \cite{BAP,BP}. But all these seem to be controlled by Planck-scales, and not susceptible to tests by current experiments. It remains a challenge to locate potential signals of Planck scale spacetime effects at presenty accessible energy scales.

\section{Acknowledgements}

It is a pleasure for Balachandran, Marmo and Martone to thank Alberto Ibort and the Universidad Carlos III de Madrid for their wonderful hospitality and support.

The work of Balachandran and Martone was supported in part by DOE under the grant number DE-FG02-85ER40231 by the Department of Science and Technology (India) and by the Institute of Mathematical Sciences, Chennai. We thank Professor T. R. Govindarajan for his very friendly hospitality at the Institute of Mathematical Sciences, Chennai. Balachandran was also supported by the Department of Science and Technology, India.


\bibliographystyle{apsrmp}

\begin{thebibliography}{99}

\bibitem{Schur1} I. Schur, {\it \"Uber die rationalen Darstellungen der allgemeinen linearen Gruppe}, Sitzungsberichte Akad. Berlin 1927, 58-75 (1927).

\bibitem{Schur2} H. Weyl, {\it The Classical Groups. Their Invariants and Representations}, Princeton University Press, Princeton, N. J. (1939).

\bibitem{Landi} G. Landi, {\it An introduction to noncommutative spaces and their geometry}, Springer (1997) [arXiv:hep-th/9701078]

\bibitem{Varilly} J. C. Varilly, {\it An Introduction to Noncommutative Geometry}, European Math. Soc. Publishing House (2006) [physics/9709045]


\bibitem{Chaichian} M. Chaichian, P. Kulish, K. Nishijima and A. Tureanu, {\it On a Lorentz-invariant interpretation of noncommutative space-time and its implications on noncommutative QFT}, Phys. Lett. B {\bf 604}, 98-102 (2004), [arXiv:hep-th/0408069].


\bibitem{Aschieri1} P. Aschieri, C. Blohmann, M. Dimitrijevic, F. Mayer, P. Schupp and J. Wess, {\it A gravity theory on noncommutative spaces}, Clas. Quant. Grav. {\bf 22}, 3511 (2005) [arXiv:hep-th/0504183].

\bibitem{Aschieri2} P. Aschieri, M. Dimitrijevic, F. Meyer, S. Schraml and J. Wess, {\it Twisted Gauge Theories}, Lett. in Math. Phys. {\bf 78}, 61-71 (2006) [arXiv:hep-th/0603024] 

\bibitem{Grosse} H. Grosse, Phys. Lett. B {\bf 86}, 267 (1979)


\bibitem{dress1} A. B. Zamolodchikov and Al. B. Zamolodchikov, {\it Factorized S-Matrices in Two Dimensions as the Exact Solutions of Certain Relativistic Quantum Field Theory Models}, Annals Phys. {\bf 120}, 253 (1979).

\bibitem{dress2} L. Faddeev, {\it Quantum completely integrable models in field theory}, Sov. Rev. C {\bf 1}, 107 (1980).


\bibitem{sasha} A. P. Balachandran, A. Pinzul and B. A. Quereshi, {\it Twisted Poincar\'e invariant quantum field theories}, Phys. Rev. D {\bf 77}, 025021(2008) [arXiv:hep-th/0708.1779]. 



\bibitem{Mario} A. P. Balachandran and M. Martone, {\it  Twisted Quantum Fields on Moyal and Wick-Voros Planes are Inequivalent} Mod. Phys. Lett. A24:1721-1730 (2009), [arXiv:hep-th/0902.1247].

\bibitem{Mario3} A. P. Balachandran, A. Ibort, G. Marmo and M. Martone, {\it  Inequivalence of QFT's on Noncommutative Spacetimes: Moyal versus Wick-Voros}, Phys. Rev. D{\bf 81}:085017 (2010), [arXiv:hep-th/0910.4779].


\bibitem{GeonMario} A. P. Balachandran, A. Ibort, G. Marmo and M. Martone, {\it Quantum Geons and Noncommutative Spacetimes} (2010).


\bibitem{Nonass} A. P. Balachandran and B. A. Qureshi {\it Poincar\'e Quasi-Hopf Symmetry and Non-Associative Spacetime Algebra from Twisted Gauge Theories}, Phys. Rev. D{\bf 81}:065006 (2010) [arXiv:hep-th/0903.0478].


\bibitem{review} A. P. Balachandran, A. Ibort, G. Marmo and M. Martone, {\it Quantum Fields on Noncommutative Spacetimes: Theory and Phenomenology}, SIGMA 6:052 (2010) [arXiv:hep-th/1003.4356].


\bibitem{BAP} A. P. Balachandran, A. Joseph and P. Padmanabhan, {\it Non-Pauli Transitions From Spacetime Noncommutativity}, Phys. Rev. Lett. {\bf 105}:051601 (2010) [arXiv:hep-th/1003.2250].

\bibitem{BP} A. P. Balachandran and P. Padmanabhan, {\it Non-Pauli Effects from Noncommutative Spacetimes} (2010) [arXiv:hep-th/1006.1185].






\end{thebibliography}

\end{document}